\documentclass[prb,twocolumn,showpacs,preprintnumbers,amsmath,amssymb,
superscriptaddress,floatfix]{revtex4}

\usepackage{graphicx,subfigure}
\usepackage{dcolumn}
\usepackage{bm}
\usepackage{psfrag}
\usepackage{multirow}

\def\be{\begin{equation}}       \def\ee{\end{equation}}
\def\bea{\begin{eqnarray}}      \def\eea{\end{eqnarray}}
\def\half{\frac{1}{2}}

\begin{document}

\title{Von Neumann entropy and on-site localization for perpetually coupled 
qubits}
\author{Kingshuk Majumdar}
\affiliation{Department of Physics, Grand Valley State University, Allendale, 
MI 49401}
\email{majumdak@gvsu.edu}
\date{\today}

\begin{abstract}\label{abstract}
We use the von Neumann entropy to study the single and many-particle on-site 
localizations of stationary states for an anisotropic Heisenberg spin-$\half$ 
chain. With a constructed bounded sequence of on-site energies for single and 
many-particle systems we demonstrate that the von Neumann entropy approaches 
zero indicating strong on-site localizations for all states. On the contrary, 
random on-site energy sequence does not lead to strong on-site confinement of 
all states. Our numerical results indicate that the von Neumann entropy provides
a new insight to analyze on-site localizations for these systems.
\end{abstract}
\maketitle

\section{\label{sec:Intro}Introduction}
Following the seminal work of P. W. Anderson in 1958~\cite{anderson1,anderson2}, 
disorder induced localization became an active area of research in condensed 
matter physics.~\cite{lic,lee} Current interest in quantum computation and 
quantum information processing has further revived the importance of studying 
localization of quantum bits or qubits.~\cite{nielson} In many proposed 
experimental realizations of a quantum computer such as ion traps~\cite{cirac}, 
nuclear magnetic resonance systems~\cite{chuang,ladd}, optical 
lattices~\cite{brennen,jaksch}, quantum dots~\cite{loss}, a key common feature 
is the presence of qubit-qubit interaction, which is needed to perform two-qubit
logical operations. The couplings of the qubits with the external environment 
lead to a finite lifetime to the excited state of a given qubit. In order to 
perform a computation an ideal quantum computer must be isolated from the outside
environment and at the same time its qubits have to be accessible to perform 
quantum gate operations. The major challenge in realizing such a quantum computer
is to obtain a delicate balance between the time for which the system remains 
quantum mechanically coherent, which is the decoherence time and the time it
takes to perform quantum gate operations.

The interactions between qubits can not
be completely turned off in a realistic quantum computer. However, it can be 
controlled so that measurements can be performed on individual qubits. Recent 
theoretical works in this direction on many-particle systems have shown a way
to achieve this.~\cite{santos,mark1,mark2} With appropriately tuning the
on-site energy levels for the qubits it has been shown that delocalization can 
be suppressed, which results in strong on-site confinement of all stationary 
states in the system. In this case the effective localization length is much 
smaller than the inter-site distance instead of an exponential decay of the 
wave-function over many sites.

Recently there has been a tremendous effort to understand the properties of 
quantum entanglement, which is one of the most intriguing features of quantum 
theory.~\cite{einstein} It plays a key role in many of the applications of 
quantum information processing.~\cite{nielson,bennett} Quantum 
entanglement measured by the von Neumann entropy has been extensively studied in
various condensed matter systems in recent years.~\cite{amico} It has been found
that the von Neumann entropy can be used to identify quantum phase transitions 
in fermionic systems~\cite{zanardi,larsson,osterloh,kopp,gu} and 
localization-delocalization transitions in interacting electron 
systems~\cite{gong,tong}.

In this paper we have explored the possibility of using the von Neumann entropy as a 
measure for identifying strong on-site localization for single and many-particle
systems of qubits. Von Neumann entropy approach provides a new insight into 
localization, which is complementary to other approaches that have been used 
before especially the inverse participation ratio (IPR).~\cite{santos,mark1,mark2}
We have considered localization for systems with constructed on-site energies
and random on-site energies. For a finite system with constructed on-site energy
sequence for qubits we show evidence for strong on-site localizations of {\em all} 
states using the von Neumann entropy. For the same system with random on-site 
energy sequence we find that it does not lead to strong on-site localization.

\section{\label{sec:model}Model}

Qubits being two level systems can be modeled as spin-$\half$ chains in a 
one-dimensional lattice in presence of an effective magnetic field along the $z$
direction. The excitation energy of a qubit will then be the Zeeman energy
of a spin and the qubit-qubit interaction will be represented by the exchange 
spin coupling. The Hamiltonian of the anisotropic spin-$\half$ chain in an effective
magnetic field is
\bea
H &=&  H_0 + H_1, \nonumber \\
H_0 &=& \sum_{n=1}^L (\epsilon_0+\epsilon_n) S_n^z + \half \sum_{n=1}^{L-1}J^z_{nn+1}S^z_nS^z_{n+1},
\nonumber \\
H_1 &=& \half \sum_{n=1}^{L-1}J_{nn+1} \left(S^x_nS^x_{n+1}+S^y_nS^y_{n+1}\right).
\label{ham}
\eea
Here $J^x_{nn+1}=J^y_{nn+1}$ and $J^z_{nn+1}$ are the exchange couplings between 
nearest neighbor sites $n,n+1$. $L$ is the number of sites and the 
parameter $\epsilon_n$ corresponds to the
Zeeman splitting of spin $n$ due to the static magnetic field in the $z$ direction.
We have considered $J^x_{nn+1}$ and $J^y_{nn+1}$
to be equal to $J$ so that the qubit interaction terms do not oscillate at 
qubit transition frequencies. Using Jordon-Wigner transformation~\cite{wigner}
the Hamiltonian for the $XXZ$ spin-chain (Eq.~\ref{ham}) can be mapped onto a 
system of spin-less interacting fermions.~\cite{fradkin} The resulting Hamiltonian, 
expressed in terms of the fermion creation and annihilation operators 
($a_n^\dagger$ and $a_n$) is
\[H = H_0 + H_1,\]
where
\bea
H_0 &=& \sum_{n=1}^L \epsilon_n a_n^\dagger a_n + 
J\Delta \sum_{n=1}^{L-1} a_n^\dagger a_{n+1}^\dagger a_{n+1} a_n, \nonumber \\
H_1 &=& \half J \sum_{n=1}^{L-1} \left(a_n^\dagger a_{n+1}+a_{n+1}^\dagger a_n \right).
\label{fermionham}
\eea
Here, $J = J^{xx}_{nn+1}\;(J > 0)$ is the fermion hopping integral, 
$J\Delta = J^{zz}_{nn+1}$ is the interaction energy of fermions on neighboring 
sites, and on-site fermion energies $\epsilon_n$ represent the on-site Zeeman 
energies of the spins (or the excitation energies of the qubits). The 
dimensionless parameter $\Delta\; (\Delta > 0)$ characterizes the anisotropy of 
the spin-spin interaction. For the on-site fermion energies $\epsilon_n$
we have used the bounded one-parameter sequence~\cite{santos} with $n \geq 1$
\be
\epsilon_n = \half h \left[ (-1)^n - \sum_{k=2}^{n+1}(-1)^{{\rm int}(n/k)}
\alpha^{k-1}\right],
\label{energyseq}
\ee
where $h$ is the inter-subband distance that significantly
exceeds the hopping integral $J$ and $\alpha \;(0\leq \alpha <1)$ is a
one-dimensional parameter. int($n/k$) represents the integer part. This 
constructed energy sequence detunes the on-site energies $\epsilon_n$ from 
others. For $\alpha=0$, $\epsilon_n$  splits into two major subbands at 
$\pm h/2$ for even and odd values of $n$. For small values of $\alpha$, the two 
major subbands have width $\approx \alpha h$ and are separated by $\approx h$. 
With increase in $\alpha$ the subbands start overlapping and finally the separation
between the subbands disappears.

To compare our results with some other on-site energy sequence we have considered a 
simple random sequence of on-site energies
\be
\epsilon_n = W r_n.
\label{randomseq}
\ee
$W$ is the band-width and $r_n$ are $n$ independent random numbers uniformly 
distributed in the interval $[0,1]$.

For our model of spin-less interacting fermions there are two possible local 
states at each site, $|1\rangle$  and $|0\rangle$ corresponding to the $n$-th 
site being occupied by a single fermion and the site being empty. The local 
density matrix at site $n$ in eigenstate $j$ is
\be
\rho^j_n = |\psi^j_n|^2(|0\rangle\langle 0|)_n + 
(1-|\psi^j_n|^2)(|1\rangle\langle 1|)_n.
\label{rhoj}
\ee
The wave-function for the N-particle eigenstate is given by,
$|\psi^j_N\rangle = \sum_{k=1}^N \psi^j_k |j\rangle$. 
In Eq.~\ref{rhoj}, $\psi_n^j$ is the amplitude of an energy eigenstate $j$ at
site $n$. The local von Neumann entropy, $S^j_n$ is defined
as~\cite{nielson}
\be
S^j_n = -|\psi^j_n|^2\log_2|\psi_n^j|^2 - 
(1-|\psi^j_n|^2)\log_2(1-|\psi_n^j|^2).
\label{entropy}
\ee
The spectrum averaged von Neumann entropy, $\langle S \rangle$ is the
entropy averaged over all the eigenstates $M$ and total
number of sites $N$
\be
\langle S \rangle = \frac 1{MN} \sum_{n=1}^N \sum_{j=1}^M S^j_n.
\label{avgentropy}
\ee
For many-particle system with $L$ sites and $N$ particles, the number of energy 
eigenstates is $M = L!/N!(L-N)!$.

One of the basic properties of the von Neumann entropy is sub-additivity. For 
two distinct quantum systems A and B the joint entropy for the two systems 
satisfy the inequality: $S(A,B) \leq S(A) + S(B)$. The equality holds if and only 
if systems A and B are uncorrelated~\cite{nielson}, which is the case with 
different realizations of random on-site energies. Thus for the random on-site 
energy sequence (Eq.~\ref{randomseq}) we have averaged the entropies 
$\langle S\rangle$ over all the different random configurations.

There are two special cases where Eq.~\ref{avgentropy} has simple analytical 
expressions. For an extended state $j$, $\psi^j_n=1/\sqrt{N}$ for all
$n$. Then it follows that the averaged entropy is a constant equal to
\be
\langle S\rangle = -\frac 1{MN}\log_2 \frac 1{N} - \frac {N-1}{MN}\log_2 
\left(1-\frac 1{N} \right).
\label{normS}
\ee
For a localized state $\psi^j_n = \delta_{nm} $ ($m$ is a given site), 
$\langle S \rangle = 0$. So the normalized $\langle S \rangle$ varies between 0 
(for localized states) and 1 (for extended states).

Another quantitative measure that is widely used to characterize localization 
is the inverse participation ratio, $I_j$ for eigenstate $j$ which is 
defined as
\be
I_j = \left( \sum_n |\psi_n^j |^4\right)^{-1}.
\ee
For our calculations we have used the inverse participation ratio averaged over 
all eigenstates $M$
\be
\langle {\rm IPR} \rangle = \frac 1{M} \sum_{j=1}^M I_j.
\label{IPR}
\ee
For fully localized states $I_j=1$. $I_j \approx 1$ is an indicator of
strong on-site localization of the $j$-th state of the system. For extended 
states $I_j >> 1$, with no upper bound.

\section{\label{sec:results}Numerical Results}

\subsection{\label{sec: singleparticle} Single Particle Localization}
We consider a single-particle system with $L = 100$ and $L = 300$ sites. For the 
on-site energies we use the constructed energy sequence in Eq.~\ref{energyseq} 
with the following parameters for the inter sub-band distance $h$ and hopping
integral $J:\; J = 1/30,h = 20J$. With $h$ much greater than $J$ the on-site 
energies are separated in two sub-bands for even and odd $n$. Next by choosing a 
dimensionless parameter $\alpha$ of the order of $J/h$ we further split each 
sub-band into two separate bands to detune the next nearest neighbors. We then 
construct the Hamiltonian in Eq.~\ref{fermionham} and diagonalize it to obtain 
the energy eigenstates. Interaction energy $\Delta$ is equal to zero for the
single-particle case. With these eigenstates we numerically calculate the 
spectrum averaged entropy $\langle S \rangle$ and the averaged inverse 
participation ratio $\langle {\rm IPR} \rangle$.
\begin{figure}[httb]
\includegraphics[width=8.5cm]{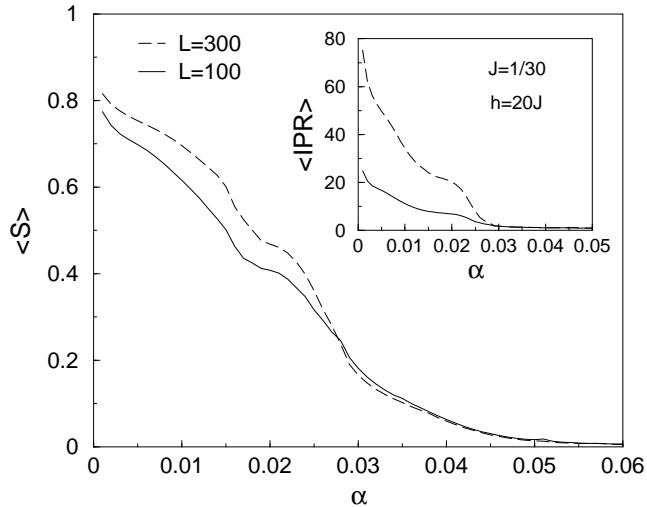}
\caption{\label{fig:single} Single-particle spectrum averaged von Neumann 
entropy $\langle S \rangle$ (normalized by the constant in Eq.~\ref{normS}) is 
plotted for $L = 100$ and $L = 300$ sites as a function of a dimensionless 
parameter $\alpha$. The constructed energy sequence (Eq.~\ref{energyseq}) is
used for the on-site energies. The hopping integral $J = 1/30$
and the inter subband distance $h = 20J$ is used for this plot.
In the inset the average inverse participation ratio, $\langle {\rm IPR} \rangle$
is plotted. Both $\langle {\rm IPR} \rangle$ and $\langle S \rangle$ decrease 
with increase in $\alpha$. Strong on-site localization 
$\langle S \rangle \rightarrow 0 $ or $\langle {\rm IPR} \rangle \rightarrow 1$ 
occurs at $\alpha \approx 0.1$ for both $L = 100$ and $L = 300$ sites chain.}
\end{figure}

In Fig.~\ref{fig:single} we plot $\langle S \rangle$ as a function of $\alpha$ for 
both $L = 100$ and $L = 300$ site chains. In the inset we show the results
for the $\langle {\rm IPR} \rangle$. For $\alpha \rightarrow 0$ the energy 
sequence in Eq.~\ref{energyseq} reduces to $\epsilon_n = (-1)^nh/2$. Hence, the 
stationary states form two bands of widths $\approx J^2/h$ (for $h >> J$) 
centered at $\pm h/2$ for even and odd $n$. With increase in $\alpha$,
the bands at $\pm h/2$ further split into more and more sub-bands. This results 
in the monotonic decrease of both the entropy and the $\langle {\rm IPR} \rangle$. 
The spatial decay of single-particle stationary states has been studied in detail 
in Ref.~\cite{santos} where it has been shown that strong on-site confinement of 
the stationary states occur as $\alpha$ exceeds a certain threshold value 
$\alpha_{\rm th} = J/2h$. Our numerical results are in good agreement with the 
theoretical estimate. We find $\langle S \rangle \rightarrow 0 $ and 
$\langle {\rm IPR} \rangle \rightarrow 1$ for $\alpha \approx 0.1$
demonstrating strong on-site localization for all states.

\begin{figure}[httb]
\includegraphics[width=8.5cm]{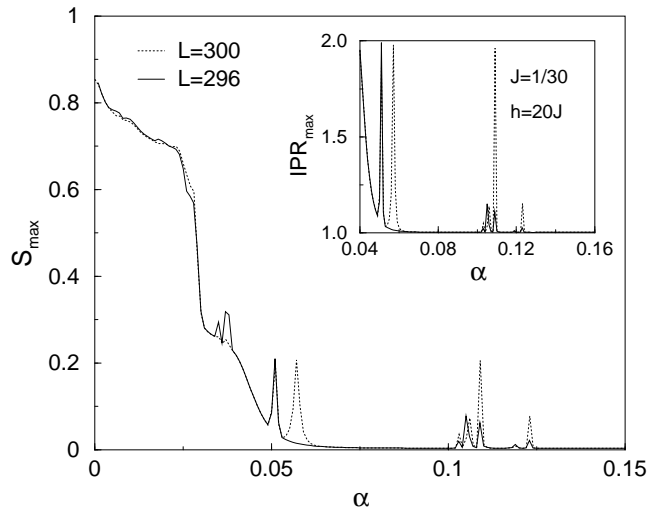}
\caption{\label{fig:single_max} Maximal single-particle von Neumann entropy, 
$S_{\rm max}$ (Maximal IPR in the inset) is plotted as function of $\alpha$ for
$L = 300$ and $L = 296$ sites. Eq.~\ref{energyseq} has been used for the energy 
sequence. Both $S_{\rm max}$ and IPR$_{\rm max}$ sharply decrease with
increase in $\alpha$ showing strong on-site single particle localization. The 
narrow peak for $L = 300$ sites chain close to $\alpha \approx 0.1$ 
is due to the boundary effect. This sharp peak greatly reduces
in size for $L = 296$ sites chain.}
\end{figure}

Fig.~\ref{fig:single_max} shows that the maximal von Neumann entropy $S_{\rm max}$ 
and maximal IPR exhibit sharp resonant peaks (shown for $L = 300$ and $L = 296$
sites). Maximal entropy for each $\alpha$ is obtained from a set of averaged 
entropies for all the eigenstates, $S_{\rm max}\equiv {\rm max}_j \langle S \rangle_j$. 
Maximal IPR is obtained similarly, ${\rm IPR}_{\rm max}\equiv {\rm max}_j I_j$. The narrow 
peaks are seen for $S_{\rm max}$ close to zero (IPR$_{\rm max}$ close to one), which 
demonstrate that very few on-sites states are hybridized with each other. The 
resonant peak near $\alpha \approx 0.1$ is due to hopping-induced shift of energy
levels of $\approx (J/2)^2/h$. For the chain with $L = 296$ sites the sharp
peak for $\alpha \approx 0.1$ is greatly reduced in size. In this case
the shift in energy is $\approx J^2/2h$. The difference of $\epsilon_n$ for
$L = 300$ and $L = 296$ sites is $\approx h\alpha^3$ for $\alpha << 1$.

\begin{figure}[httb]
\includegraphics[width=8.5cm]{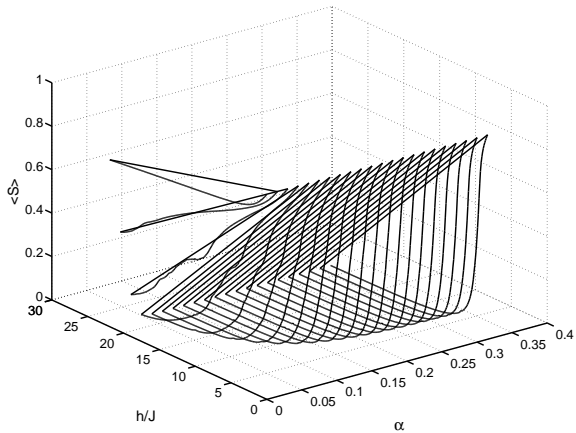}
\caption{\label{fig:single_3D} Three dimensional plot of the single particle von
Neumann entropy, $\langle S\rangle$ as a function of a $\alpha$ and $h/J$ for
$L = 100$ sites and for the energy sequence. The entropy sharply decreases with 
increase in $\alpha$ and $h/J$. The minimum $\langle S\rangle = 0.0031$ occurs 
for $\alpha = 0.166$ and $h/J = 22$.}
\end{figure}

The rationale for choosing the different parameters in
the model is shown in Fig.~\ref{fig:single_3D}, where we have plotted the
von Neumann entropy as a function of $\alpha$  and $J/h$ for the
single-particle system with $L = 100$ sites. The entropy
decreases sharply with increase in $\alpha$ and $h/J$ reaching a
minimum value $\langle S\rangle = 0.0031$ for $\alpha = 0.166$ and 
$h/J = 22$. So for small $\alpha$ and $h/J \approx 20$ {\em most of the states} 
are locally confined on sites.

\begin{figure}[httb]
\includegraphics[width=8.5cm]{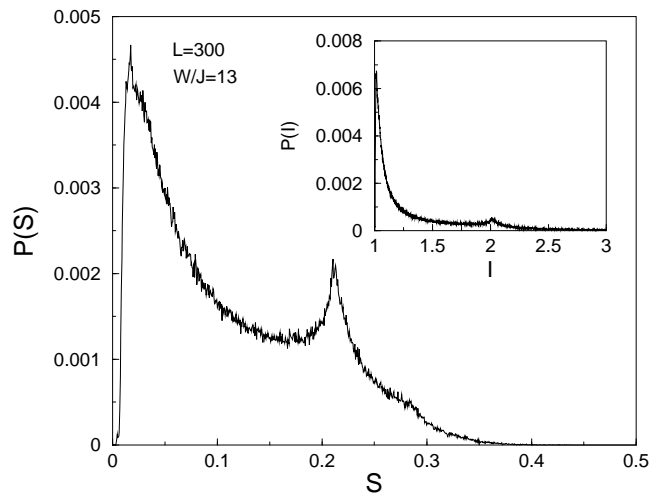}
\caption{\label{fig:single_random} Normalized single-particle probability 
distribution of the spectrum averaged von Neumann entropy, $P(S)$ is 
plotted for $L = 300$ sites with random energy sequence (Eq.~\ref{randomseq}) of
bandwidth $W = 13J$. The entropy has been averaged over an ensemble of 2000 
random on-site energy configurations. In the inset the probability distribution 
for the inverse participation ratio, $P(I)$ is shown. The sharp peak at 
$S \approx 0$ (and $I \approx 1$ in the inset) is the indication of localization. 
However, the distributions are broad indicating strong hybridization between
many sites. Multiple resonances occur due to these hybridization one of which we 
find for $S \approx 0.2$ or $I \approx 2$.}
\end{figure}

In Fig.~\ref{fig:single_random} we have plotted the normalized probability distribution 
$P(S)$ and $P(I)$ (in the inset) for the random on-site energy sequence 
(Eq.~\ref{randomseq}). 
The probability distributions (histogram plots) are obtained by averaging 
the entropy and inverse participation ratio over 2000 random on-site energy 
configurations. Sharp peak close to $S \rightarrow 0$ or $I \rightarrow 1$ 
demonstrate that many of the states are localized. However, the distributions 
are broad and resonances are seen indicating strong hybridization between 
neighboring sites, which means {\em all the states} are not confined.

The entropy and 
the IPR for the constructed energy sequence differ significantly from the 
random on-site energy sequence. For the constructed sequence we find strong
on-site localization of all states whereas with the random sequence the 
confinement is not so strong.

\subsection{\label{sec: manyparticle} Many Particle Localization}
In this section we investigate the on-site localization of all states for a 
many-particle system with $L = 12$ sites and $N = 6$ particles where no two 
particles can occupy the same site. Here we have a total number of $12!/6!6! = 924$ 
ways of putting six particles in twelve sites. We construct the Hamiltonian in 
occupation number basis states and then diagonalize the $924 \times 924$ 
Hamiltonian matrix to obtain the energy eigenstates.
\begin{figure}[httb]
\includegraphics[width=8.5cm]{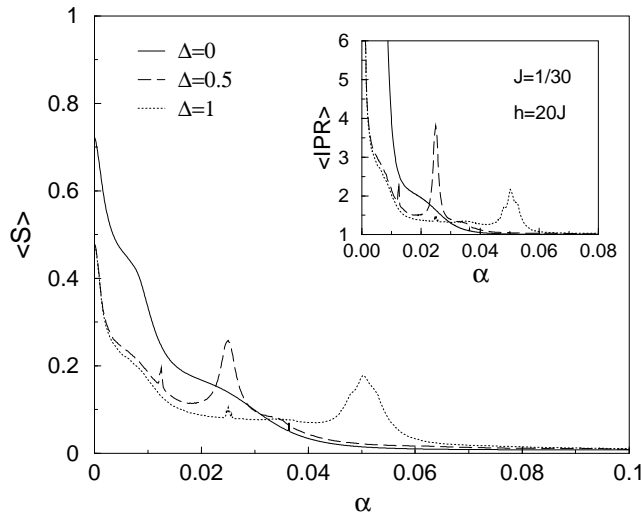}
\caption{\label{fig:many} Spectrum averaged von Neumann entropy for a system 
with $N = 6$ particles in $L = 12$ sites and with constructed energy sequence 
plotted against $\alpha$. Three values of the interaction parameter 
$\Delta=0,0.5,1$ are used for the plots. In the inset averaged IPR is plotted. 
Strong on-site localization $\langle S \rangle \rightarrow 0$ or 
$\langle {\rm IPR} \rangle \rightarrow 1$ occurs for $\alpha$ close to 0.1 for 
all three values of $\Delta$. For $\Delta \neq 0$ isolated peaks at various 
values of $\alpha$ is due to hybridization of resonating on-site many-particle
states.}
\end{figure}

\begin{figure}[httb]
\includegraphics[width=8.5cm]{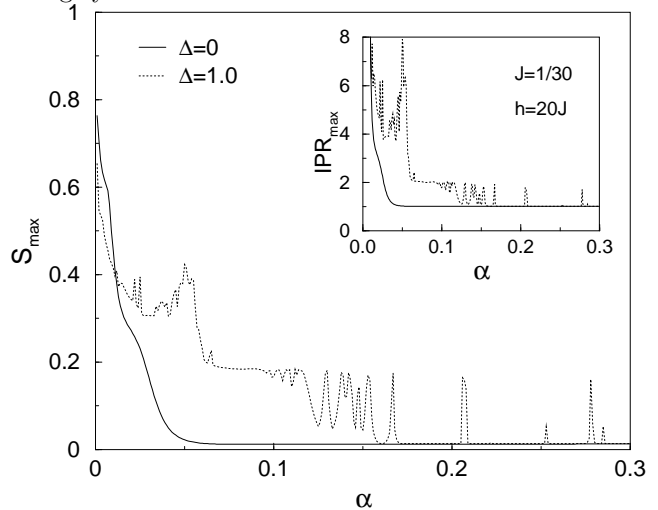}
\caption{\label{fig:many_max} Plot of maximal many-particle von Neumann entropy
(maximal IPR in the inset) for $N = 6$ particles in $L = 12$ sites. Two values 
of the interaction parameter $\Delta=0,1$ are used. Both $S_{\rm max}$ and 
IPR$_{\rm max}$ sharply decrease with increase in $\alpha$ showing strong on-site 
single particle localization. However, sharp isolated peaks are seen for 
$\Delta=1.0$, which indicates that some of the stationary states are not 
localized on-site. These states are strongly hybridized with other resonating 
on-site many-particle states.}
\end{figure}

In Fig.~\ref{fig:many} we have plotted the entropy (and IPR in the inset) as a function 
of the dimensionless parameter $\alpha$ for the on-site energy sequence in 
Eq.~\ref{energyseq}. The entropy and the IPR have been averaged over all the 924 
eigenstates. Three different values of the interaction parameter $\Delta= 0,0.5$,
and 1 are chosen for the plots. $\Delta =0$ corresponds to only $XX$ type coupling 
between the nearest neighbor spins -- this case is similar to the single-particle 
case. The decrease in $\langle S \rangle$ with increase in $\alpha$ is seen for 
all three values of $\Delta$. Fig.~\ref{fig:many_max} shows the decrease in the 
maximal entropy and maximal IPR (in the inset) with increase in $\alpha$ for 
$\Delta=0$ and 1.

Localization of many-particle system differ from single-particle system due to 
the interaction term $J\Delta$ in the Hamiltonian. Fig.~\ref{fig:many} shows 
that the entropy (and IPR) decreases with increase in $\Delta$. For 
$\Delta >> J/h$, the energy bands at $\pm h/2$ split into subbands. Such splitting 
lowers the value of $\langle S \rangle$ and IPR for non-zero $\Delta$.

For non-zero $\Delta$, many particle resonances are seen in Fig.~\ref{fig:many} 
and Fig.~\ref{fig:many_max} for small values of $\alpha$, indicating some
stationary states are no longer on-site localized. Strong on-site localization 
of the states require that the energies of the distant sites must be tuned away 
from each other. This happens for large values of $\alpha$, where we find 
$\langle S \rangle$ approaches zero as $\alpha \rightarrow 0.1$ demonstrating 
strong on-site localization of all states.

Similar to the single-particle case, we have calculated the three-dimensional 
plot of entropy as a function of $\alpha$ and $h/J$. We find that the von Neumann 
entropy decreases with increase in $\alpha$ and $h/J$. The minimum value
of $\langle S \rangle = 0.0061$ occurs for $\alpha \approx 0.23$ and 
$h/J \approx 22$.
\begin{figure}[httb]
\includegraphics[width=8.5cm]{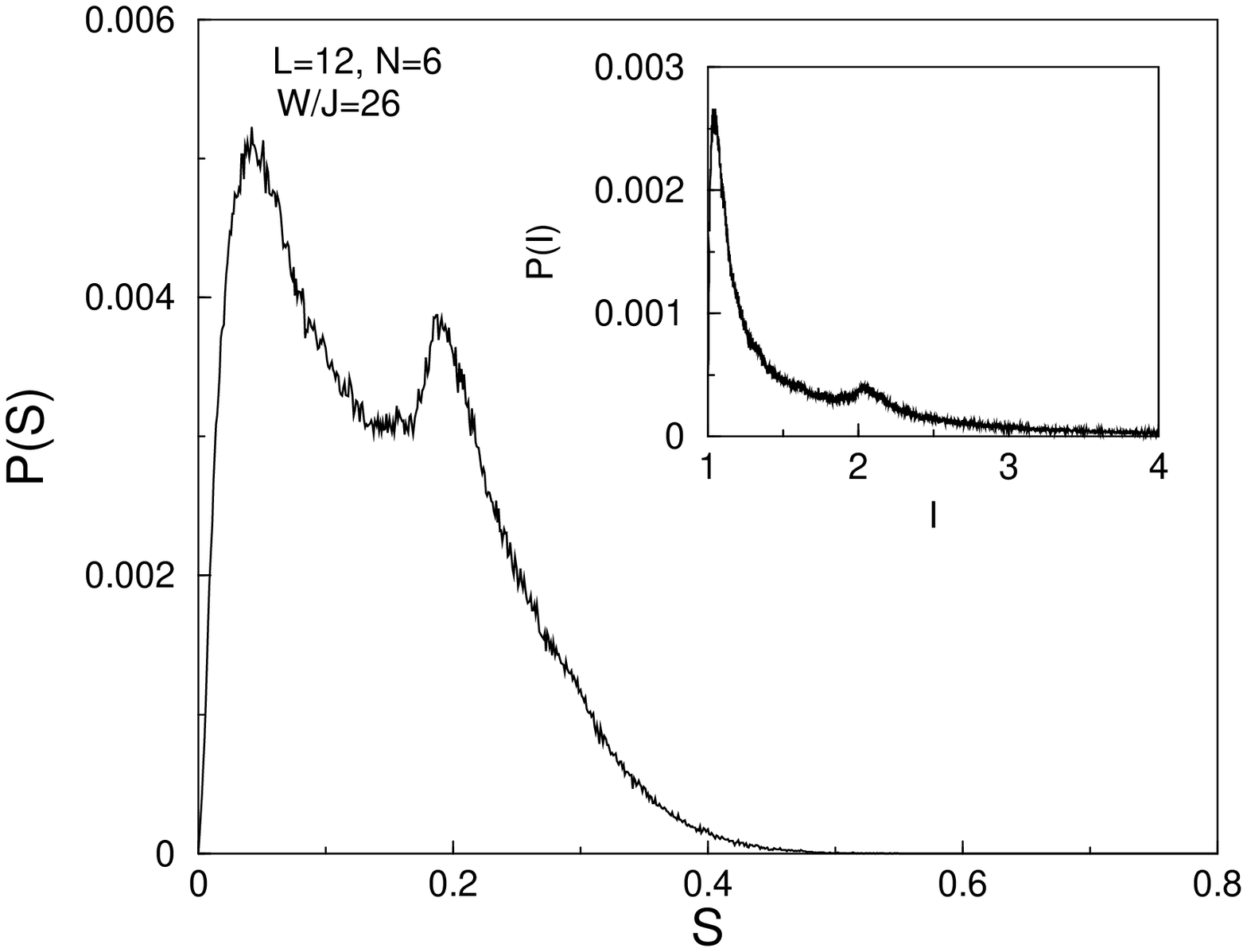}
\caption{\label{fig:many_random} Normalized probability distribution of the 
von Neumann entropy, $P(S)$ is plotted for $N = 6$ particles in $L = 12$
sites with random energy sequence. Band-width is chosen to be $W = 26J$. The 
entropy has been averaged over an ensemble of 2000 random on-site energy 
configurations. In the inset the probability distribution for the inverse 
participation ratio, $P(I)$ is shown. The sharp peak at $S \approx 0$ (and 
$I \approx 1$ in the inset) is the indication of localization. However, the 
distributions are broad indicating strong hybridization between many sites. 
Peaks at $S = 0.2$ or $I = 2$ occur due to hybridization between many particle 
states.}
\end{figure}

Fig.~\ref{fig:many_random} shows the normalized probability distribution for
the von Neumann entropy, $P(S)$ (normalized probability distribution for the 
inverse participation ratio, $P(I)$ in the inset) for the random on-site energy 
sequence. As before both the entropy and the IPR have been averaged over 2000 
random on-site energy configurations. Peak at $S \approx 0$ shows localization 
of many of the states but not all. The distribution is broad and many particle 
resonances are seen (as an example the smaller peak at $S \approx 0.2$),
which means that many on-site states are strongly hybridized as the stationary 
wave-functions spread over many sites. This is in contrast to the case with 
constructed energy sequence where {\em most} of the states are strongly confined 
for $\alpha > 0.1$.

\section{Conclusions}
In this paper we have explored the use of von Neumann entropy to characterize strong 
on-site localization for interacting qubits. For a successful realization of 
a quantum computer strong on-site localization of all states is a necessity. 
We have considered both single and many-particle systems with a carefully 
constructed sequence of on-site energies. With our numerical results we have 
demonstrated that von Neumann entropy approaches zero for such systems with 
certain values of the parameters in the model, thus indicating strong on-site 
localization of {\em all states}. For comparison we have studied single and 
many-particle systems with random on-site energies where we find hybridization 
between many sites. All the sites for the random case are not confined which 
results in a wide probability distribution of the entropy with multiple 
resonance peaks.

In conclusion, we have found that the von Neumann entropy provides a systematic 
approach to the problem and is thus an effective tool to study 
localization-delocalization transitions in these many particle interacting 
systems.

\section{Acknowledgments}
The author gratefully acknowledges helpful discussions with M. I. Dykman. 
This work was supported by a grant from the Research Corporation.

\bibliography{qubit}
\end{document}